\documentclass[11pt,hyper,letter]{JHEP3}
\usepackage{graphicx,amssymb,amsmath,amsfonts,cite}
\usepackage[english]{babel}
\usepackage{amssymb}
\usepackage{amsfonts}
\usepackage{amsmath}
\usepackage{graphicx}
\usepackage{epsfig}
\usepackage{cite}
\newcommand{\bea}{\begin{eqnarray}}
\newcommand{\eea}{\end{eqnarray}}
\newcommand{\non}{\nonumber \\}
\newcommand{\CR}{\non\cr}
\newcommand{\pa}{\partial}

\newcommand\be{\begin{equation}}
\newcommand\ee{\end{equation}}

\vspace{18pt}
\title   {\Large
     Large N Chern-Simons with massive fundamental fermions - \\
     A model with no bound states}
\author
   { Yitzhak Frishman${}^1$\footnotemark[1]  and Jacob Sonnenschein${}^2$\footnotemark[2] \\
${}^1$\textit {
     Department of Particle Physics and Astrophysics\\
The Weizmann Institute of Science, Rehovot 76100, Israel} \\
\\
${}^2$\textit{
 	 The Raymond and Beverly Sackler School of Physics and Astronomy,\\
	Tel Aviv University, Ramat Aviv 69978, Israel}\\

\footnotetext[1]{E-mail address: \email{yitzhak.frishman@weizmann.ac.il}}
\footnotetext[2]{E-mail address: \email{cobi@post.tau.ac.il}}

}
\vspace{18pt}
\abstract{
In a previous paper \cite{Frishman:2013dvg}, we analyzed the theory of massive fermions in the fundamental representation coupled to a $U(N)$  Chern-Simons gauge theory in three dimensions  at level $K$. It was done  in the large $N$, large $K$ limits where $\lambda=\frac{N}{K}$ was kept fixed. Among other results, we showed there that there are no high mass ``quark anti-quark" bound states. Here we show that there are no bound states at all.}


\preprint{WIS/14/08-SEP-DPPA,  TAUP-2987/14}
\newpage

\begin{document}

%%%%%%%%%%%%%%%%%%%%%%%%%%%%%%%%%%%%%%%%%%%%%%%%%%%%%%%%%%%%%%%%%%%%%%%%%%%%%%%%%%%%
%\tableofcontents

\section {Introduction}\label{Intoduction}
In recent years a  major progress has been made in the understanding  of large $N$ three dimensional  Chern-Simons theory coupled to matter in the fundamental representation \cite{Giombi:2011kc}-- \cite{Banerjee:2012gh}. Interesting exact results have been derived without the aid of supersymmetry. Among these achievements is the determination of   the exact planar free energy of the theory at finite temperature on ${\cal R}^2$ as a  function of the 't Hooft coupling $\lambda=\frac{N}{K}$, where $K$ is the level of the Chern-Simons
term. Another property of these theories is the fact that classically in the large $N$ there is an infinite tower of high-spin  conserved currents. It was shown in \cite{Giombi:2011kc} that the divergence of these currents is  equal to double  and triple trace  of currents, that vanish in the large N limit, as the former is multiplied by $1\over K$ and the latter by $1\over K^2$. In \cite{Aharony:2011jz} it was shown that in the large $N$ limit the theory of $N$ scalars coupled to $U(N)$ CS theory at level $K$ is equivalent to the Legendre transform of the theory of $K$ fermions coupled to a $U(K)$ CS theory at level $N$.

%{\bf add more about the recent developements}

In \cite{Giombi:2011kc} the fact that one can extract exact results is attributed to the discrete nature of the CS  coupling constant, the large $N$ limit, the light-cone gauge and the fact that for the massless case the theory is conformal invariant.
The main question addressed in our previous work \cite{Frishman:2013dvg} was  to what extent can one decipher  the large N CS theory coupled to massive fundamental fermions. Thus our question was essentially whether two of the three ingredients of the CS coupling, large $N$ and the light-cone gauge are enough to enable us to solve it exactly or is conformal symmetry necessary for that.  Our answer was that there  are interesting physical quantities  that can be determined even without conformal invariance.
 Concretely we had addressed the following  three questions: (i) The fermion propagator and the thermal free energy. (ii) The hight spin currents and their classical conservations. (iii) The spectrum bound state mesons.

Following \cite{Giombi:2011kc} we showed that  by solving a Schwinger-Dyson equation, the fermion propagator and the partition function at finite temperature  can be determined exactly. We generalized the result of \cite{Giombi:2011kc} to the massive case while  using a  somewhat different technique. In \cite{Aharony:2012ns}  it was shown that the result of \cite{Giombi:2011kc} is incomplete and that   there is an additional contribution to the thermal  free energy from   winding modes. The full expression written down in that paper  holds for fermions of any mass,  with an appropriate modification of the parameters.

We proved  that in the large $N$ limit  there exists an infinite set of  classically conserved high spin currents.
The conservation holds classically for high spin currents which are similar to the ones used in the massless case apart from the following replacement
\be
\label{add mass}
(\overleftarrow{D_{\sigma}}\overrightarrow{D^{\sigma}})\rightarrow (\overleftarrow{D_{\sigma}}\overrightarrow{D^{\sigma}})-m^2
\ee
 The divergence of these currents is equal to double and triple trace operators, which vanish in the large $N$ limit.  This is the same structure as for the conformal invariant setup.

As for the  spectrum of bound state mesons, we wrote down,  in analogy to the seminal work of 't Hooft on two dimensional QCD,  a Bethe-Salpeter equation for the wave function of a ``quark anti-quark" bound state. We showed that unlike the two dimensional QCD case, the three dimensional Chern-Simons theory does not admit a confining spectrum. In fact, no high mass bound states exist.

In this paper we extend the latter fact, to show that there are no bound states altogether.

%{\bf add more about our approximations}

The paper is organized as follows: The next section describes  the basic setup of a Chern-Simons theory in Euclidean three dimensions in the large $N$ and large level
 $K$ limits with fixed ratio, coupled to a fermion in the fundamental representation. Section \S 3 is devoted to the determination of the fermion propagator at zero temperature.
In section \S 4 we write down a 't Hooft-like equation for the bound states of the theory at zero temperature, and transform it to a form closer to the two dimensional case. We then show in section \S 5  that there are no bound states solutions.

In \S 6 we review the connection to the bosonic theory, showing that the correspondence implies no bound states in the fermionic case, as we found. In \S 7we compare
with \cite{Bardeen:2014qua}, where it was claimed that bound states do exist.

In the last section we summarize our results and present several open questions.

\section{The setup}\label{Ts}
The ${\cal R}^3$  Euclidean action of the $U(N)$ CS theory coupled to a massive  fermion in the fundamental representation is
\be\label{actionfermion}
S=\frac{iK}{4\pi}\int d^3x Tr [A d A + \frac{2}{3} A^3]  + \int d^3x\bar\psi( \gamma^\mu D_\mu + m_{bare}) \psi
\ee
where $A= A^a T^a$, $T^a$ is a fundamental generator normalized so that $Tr[(T^a)^2]=\frac12$ and $D_\mu\psi= \pa_\mu\psi - iA_\mu^aT^a\psi$. Note that we set the gauge coupling constant  to one.
Using light-cone coordinates $x^+,x^-, x^3$ and light-front gauge $A_-=0$ the action in
 momentum space  reads
\bea
S&=& \int\frac{ d^3 p}{(2\pi)^3}\left [ - \frac{iK}{2\pi}Tr[A_3(-p)p_- A_+(p)] +\bar\psi(-p)( i\gamma^\mu p_\mu + m_{bare}) \psi(p)\right ]\CR
&-& i\int\frac{ d^3 p}{(2\pi)^3}\int\frac{ d^3 q}{(2\pi)^3}\left [\bar\psi(-p)[\gamma^+A_+(-q) + \gamma^3 A_3(-q)]\psi(p+q)\right ]\CR
\eea

Here
\be
x^{\pm}={1\over{\sqrt 2}} [x^1 \pm x^2]\qquad
A^{\pm}={1\over{\sqrt 2}} [A^1 \pm A^2]
\ee

It follows from this action that the gauge field propagator takes the form
\be
<A^a_\mu(p) A^b_\nu(-q)> = (2\pi)^3\delta (p-q)\delta^{ab}G_{\mu\nu}(p)
\ee
where the only non-trivial components of $ G_{\mu\nu}(p)  $ are
\be
G_{+3}(p)=-G_{3+}(p)= \frac{4\pi i}{K} \frac{1}{p^+}
\ee
This translates in configuration space to
\be
<A_3(x)A_+(0)>= -<A_+(x)A_3(0)>= \frac{2}{K} \frac{\delta(x_3)}{x^+}
\ee
\section{The fermion propagator}\label{Tfp}

\begin{figure}[h]
\begin{center}
\vspace{3ex}
\includegraphics[width= 100mm]{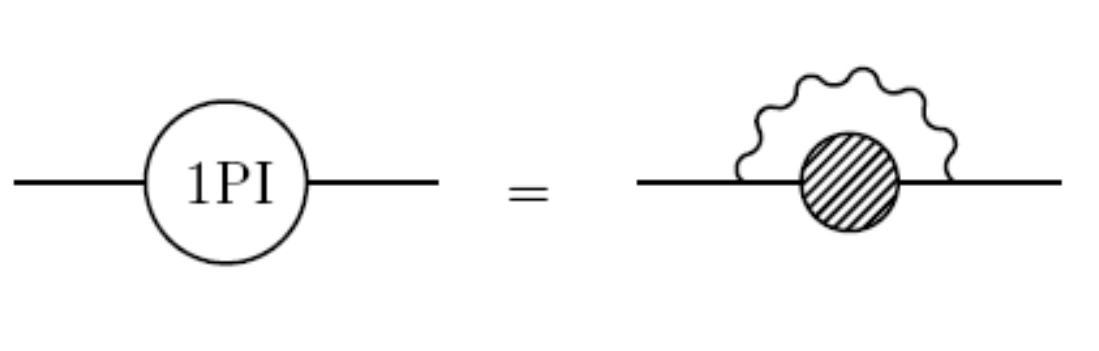}
\end{center}
\caption{Fermion Self Energy}
\label{Self}
\end{figure}

The fermion propagator is given by
\be
<\psi^m(q)\psi_n(-p)>= (2\pi)^3\delta(q-p)\delta^{m}_n S(q) = (2\pi)^3\delta(q-p)\delta^m_n \frac{1}{i\gamma^\mu q_\mu + m_{bare} + \Sigma(q)}
\ee
 where $m,n=1,... N$ and
\be\label{sigmaexpand}
\Sigma(q)= i\Sigma_\mu \gamma^\mu +\Sigma_I I  - m_{bare} I
\ee
The equation for $\Sigma$ takes the form
\be\label{FP}
\Sigma(p) = -i 4\pi \lambda\int\frac{ d^3 q}{(2\pi)^3}\frac{\gamma^+ \Sigma_I+ i\emph{I}(q+\Sigma(q))_-}{(q_\mu+\Sigma_\mu(q))(q^\mu+\Sigma^\mu(q)) + \Sigma_\emph{I}(q)^2}\frac{1}{(p-q)^+}
\ee
This is depicted in fig.(\ref{Self}).

Equating the coefficients of the various $\gamma^\mu$ matrices it is clear that
$\Sigma$ is independent of $p_3$ and
\be\label{defgf}
\Sigma_-=\Sigma_3=0 \qquad  \Sigma_I=p_sf_0(\lambda,p_s,m_{bare})\qquad \Sigma_+=p_+g_0(\lambda,p_s,m_{bare})
\ee
with
\be
p_s = \sqrt{p_1^2+p_2^2}= \sqrt{2}|p^-|=\sqrt{2}|p^+|.\ee
Substituting \ref{defgf} into (\ref{FP}) we get the following integral equations for $f_0$ and   $g_0$
\bea\label{g0f0}
g_0  &=& -\frac{4\pi\lambda}{p^-}\int\frac {d^3 q}{(2\pi)^3}\frac{q_s f_0}{q_3^2 + q_s^2(1+ g_0+ f_0^2)}\frac{1}{(p^+-q^+)}\CR
f_0 p_s - m_{bare}  &=& {4\pi\lambda}\int\frac {d^3 q}{(2\pi)^3}\frac{q^+}{q_3^2 + q_s^2(1+ g_0+ f_0^2)}\frac{1}{(p^+-q^+)}\CR
\eea
To solve for the functions $f_0(\lambda,p_s,m_{bare})$ and $g_0(\lambda,p_s,m_{bare})$ we now employ the identity
\be\label{delta}
\frac{\pa}{\pa p^-} \frac{1}{p^+} = 2\pi \delta^2 (p)
\ee
Applying this to Eq (\ref{g0f0}) we get
\bea\label{DE}
\frac{\pa}{\pa p^-} (p_s f_0) &=& \frac {\lambda p^+}{p_s \sqrt{1 + g_0 + f_0^2}} \CR
\frac{\pa}{\pa p^-} (p^- g_0) &=& - \frac { \lambda f_0}{\sqrt{1 + g_0 + f_0^2}}
\eea

Multiplying the first with $f_0 p_s$, the second by $p^+$ and adding, we get zero for the right hand side,
thus obtaining

\be
(1+ \frac{1}{2} p_s \frac {\pa}{\pa p_s})(g_0 + f_0^2) = 0
\ee

Which gives the solution

\be\label{fgrelation}
g_0 + f_0^2 = \frac {m^2}{p_s^2}
\ee

The constant of integration comes out to be $m^2$, where $m$ is the pole in the full propagator.
Using this, the first equation in (\ref{DE})  can be integrated, to give
\be
p_s f_0 = \lambda \sqrt {p_s^2 + m^2} + C
\ee
To determine C, we will evaluate the integral in (\ref{FP}) for $p_s=0$. Actually, it is enough to
evaluate the scalar part. So we have
\be
p_s\rightarrow 0 :(p_s f_0- m_{bare}) \rightarrow  -4\pi \lambda\int\frac{ d^3 q}{(2\pi)^3}\frac{1}{q^2 + m^2}
\ee

The integral is equal to
\be
-\lambda \frac{2}{\pi} \int_0^{\infty} dq + \lambda m
\ee

Setting the linearly divergent integral to zero, by dimensional regularization, we get that $C=m_{bare}$,
and so
\be\label{initial f}
p_s f_0 = \lambda \sqrt{p_s^2 + m^2} + m_{bare}
\ee

As for $p_+ g_0$, it follows from the integral equation (\ref{FP}) that it vanishes at $p_s = 0$.
This means that $p_s f_0$ equals $m$ for  $p_s = 0$, entailing
\be\label{mbare}
m_{bare} =  m (1-\lambda)
\ee
Thus the functions  $f_0$ and $g_0$ and hence  the non-trivial components of $\Sigma$  are given by
\bea\label{fg}
p_sf_0(\lambda,p_s,m_{bare})&=& m +\lambda[\sqrt{p_s^2 + m^2} -m] \CR
p_s^2g_0(\lambda,p_s,m_{bare}) &=& -\lambda \left[2m(1-\lambda)[\sqrt{p_s^2 + m^2}-m]+\lambda p_s^2 \right]
\eea

Note that we got this solution without solving for the integrals, just by their form and their values at $p=0$.

It follows from (\ref{fg}) that $\Sigma$ takes the form
\be
\Sigma(p)= ip_+\left [ -\lambda^2 -2\lambda(\frac{m_{bare}}{p_s})(\sqrt{1+\frac{m^2}{p_s^2}}- \frac{m}{p_s}) \right ]\gamma^+ +\lambda p_s\sqrt{1+\frac{m^2}{p_s^2}} I
\ee
Thus the coefficient of the unit matrix in $\Sigma$, which was  for  the massless case $\lambda p_s$, is still linear in $\lambda$ but there is a  re-scaling of $p_s\rightarrow p_s\sqrt{1+\frac{m^2}{p_s^2}}$. The coefficient of  $\gamma^+$ , $i p_+g_0$, which for the massless case was $-ip_+\lambda^2$, is determined  in the massive case  from the relation (\ref{fgrelation}).

It is easy to check that for the massless  limit these results  go back to
\be
f_0= \lambda \qquad g_0 = -\lambda^2
\ee
To summarize the propagator of the massive fermion takes the form
\be
S(q) = -\frac{iq_-\gamma^-+ iq_+(1-g_0)\gamma^+ +iq_3\gamma^3 -f_0 q_s}{q^2+m^2}
\ee

%%%%%%%%%%%%%%%%%%%%%%%%%%%%%%%%%%%%%%%%%%%%%%%%%%%%%%%%%%%%%%%%%%%%%%%%%%%%%%%%%%%%%%%
\section{'t Hooft like equation for the spectrum of bound states}\label{tHlwftsobs}
In the conformal setup when the fermions are massless a natural question to address is the spectrum of dimensions of the   primary operators  and  their descendants. The primaries are the operator $\bar \psi \psi$ and the tower of symmetric traceless currents $J^{(s)}_{\mu_1,...\mu_s}$, which are   constructed  from a fermion  anti-fermion bilinear sandwiching  derivatives and a gamma matrix.  The analysis of the spectrum of dimensions was carried out  in \cite{Giombi:2011kc}. The analogous question for the massive theory is the mass spectrum of bound states. The latter can be built in the same way as in the conformal theory.  Here we will discuss a special class of the mesonic bound states.   Note also that since the theory is invariant under  local  $U(N)$ symmetry, and not only $SU(N)$, baryon bound states are not gauge invariant
\footnote {Note, however, that at large $N$ the $U(1)$ part is down by $1\over N$.}.
We address the question of the spectrum of masses only at zero temperature.

\begin{figure}[h]
\begin{center}
\vspace{3ex}
\includegraphics[width= 100mm]{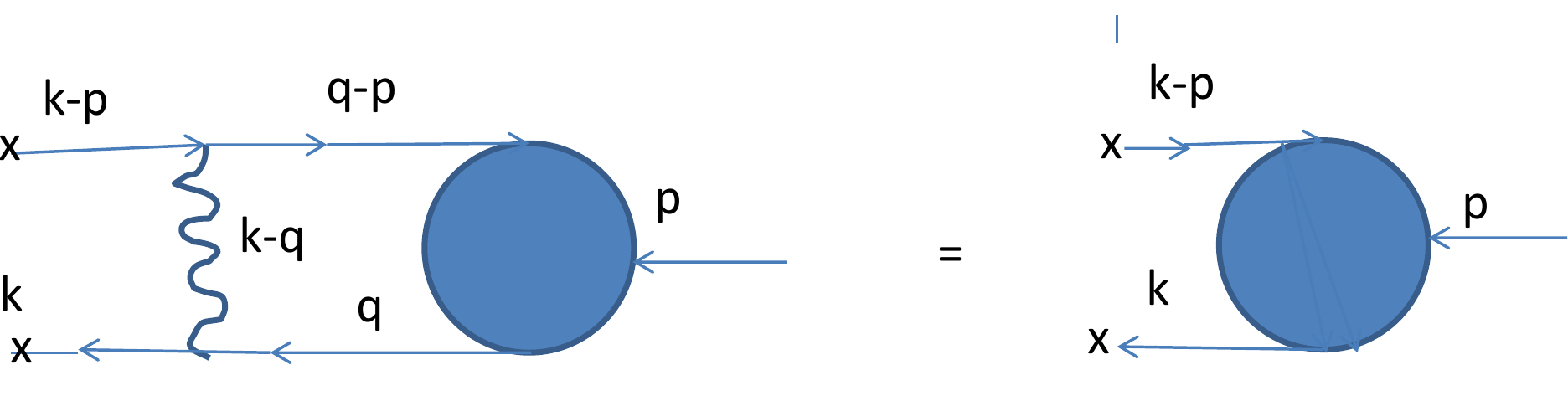}
\end{center}
\caption{Schwinger-Dyson equation for the fermion anti-fermion bound state}
\label{BethSal}
\end{figure}

The spectrum of fermion anti-fermion  bound states of two dimensional QCD in the planar limit was solved in the seminal work of 't Hooft  \cite{'tHooft:1974hx}\cite{FSbook}. Since like in that work, here we are also using (i) light-front coordinates, (ii) light-cone gauge and (iii) the planar limit, it calls for the use of a similar approach to the one used in \cite{'tHooft:1974hx}  for our system. The key player is the bound state ``wave-function" or the ``blob" which is the Fourier transform of the matrix element of the operator $\psi(x) \bar \psi(0)$ between the vacuum and the meson states,
\be\label{phi}
\phi(k, p)= \int \frac{d^3x}{(2\pi)^3}e^{ikx} <meson(p)|T \psi(x) \bar \psi(0)|0>
\ee
To determine the ``wave-function" one has to solve  a  Bethe-Salpeter   which is depicted in figure  (\ref{BethSal}).

For gauge invariance we need also a factor of $e^{i\int_0^x A_\mu(y) dy^\mu}$ between $\psi$ and $\bar\psi$. In the gauge $A_-=0$, only the integrals over  $A_+ dy^+$ and $A_3 dy^3$ will appear. As we will integrate over $k_3$ (see following (\ref{inteq})), this means $x^3=0$, leaving us with only  an integral over $A_+ dy^+$.
The discussion after eqn. (\ref{intequk3}) indicates that actually $x^+=0$. Thus we discuss here directly the matrix element (\ref{phi}).

 Note that the correlator in the definition of $\phi(p,k)$ includes the operators $\psi$ and $\bar \psi$  at different points $\psi(x) \bar \psi(0)$. Expanding $\psi(x)$ around $x=0$ we get bilinear operators of $\psi$ and $\bar \psi$ with any arbitrary number of derivatives
$(\pa_{\mu_1}...\pa_{\mu_n}\psi)  \bar \psi |_{x=0}$. Thus the blob describes  a bound state of a quark and an anti-quark with all possible orbital momenta. As for the internal spin, $\phi(p,k)$ is a 2x2 matrix, so it includes the spin zero and one components, and those are all the Dirac bilinear combinations in 3 dimensions.

To determine the masses of the bound states, we will have to go back to Minkowski space.
But let us first continue in Euclidean space.

The integral equation reads
\be\label{Int1}
 \phi(k, p)= -\frac{\lambda}{2\pi^2} S(k) \int\frac{d^3q}{(k^+ -q^+)} \left[ \gamma^+ \phi(q,p) \gamma^3 - \gamma^3 \phi(q,p) \gamma^+ \right ] S(k-p)
\ee

Using  (\ref{delta})  we can transform the integral equation into the following  differential equation
\be
\frac{\pa}{\pa k^-} \left[ S^{-1}(k) \phi(k,p) S^{-1}(k-p) \right] = \frac{\lambda}{\pi} \left[ \gamma^3\tilde\phi(\tilde k,p)\gamma^+ - \gamma^+ \tilde\phi(\tilde k,p) \gamma^3 \right ]
\ee
where $\tilde k$ is the vector $(k_1,k_2)$, and
\be
\tilde \phi(\tilde k,p) = \int d k^3\phi(p,k)
\ee

Next we expand the blob in terms of the coefficients of $\gamma^\mu$ and $\emph{I} $  similar  to (\ref{sigmaexpand})
\be
\phi= \phi_-\gamma^- + \phi_+\gamma^+ + \phi_3\gamma^3 + \phi_s \emph{I}
\ee
which gives
\be
\gamma^3 \phi \gamma^+ - \gamma^+ \phi \gamma^3 = 2[\phi_s \gamma^+ - \phi_-\emph{I}]
\ee
and similarly for $\tilde\phi$.

Thus the right hand side of (\ref{Int1}) involves $\phi_s$ and $\phi_-$ only.
This results in two coupled integral equations for $\phi_s$ and $\phi_-$, with $\phi_+$ and $\phi_3$ determined from $\phi_s$ and $\phi_-$.

The  integral equation now implies
\bea\label{ineqfsfo}
\frac{\pi^2}{\lambda}\phi_s (k,p) &=& [a_-(b_s-b_3)+ b_-(a_3+a_s)]\int\frac{d^3q}{(k^+ -q^+)}\phi_s(q,p)\CR
&-&  [a_-b_++a_+b_- +a_3b_3+ a_s b_s]\int\frac{d^3q}{(k^+ -q^+)}\phi_-(q,p)\CR
\frac{\pi^2}{\lambda}\phi_- (k,p) &=&[2 a_-b_-]\int\frac{d^3q}{(k^+ -q^+)}\phi_s(q,p)\CR
&-&  [ a_-(b_3 + b_s)+b_-(a_s-a_3)]\int\frac{d^3q}{(k^+ -q^+)}\phi_-(q,p)\CR\eea
where
\be
a_s =-\frac{f_0 k_s}{k^2+m^2} \qquad a_-=\frac{ik_-}{k^2+m^2}\qquad a_+ =\frac{ik_+(1-g_0)}{k^2+m^2}\qquad a_3= \frac{ik_3}{k^2+m^2}
\ee
and $(b_s,b_-, b_+, b_3)$ are given similarly with the same expressions but with $k-p$ replacing $k$.

Let us choose now the frame
\be
p = (0, 0, p_3)
\ee

Then
\bea
 a_- &=& \frac{i k_-}{[k^2 + m^2]}  \qquad \ \ \ \ \ \ \
 b_- = \frac{i k_-}{[k_s^2 + (k_3 - p_3)^2 + m^2]}  \CR
 a_+ &=& \frac{i k_+[1 - g_0(k_s)]}{[k^2 + m^2]}  \qquad
 b_+ = \frac{i k_+[1 - g_0(k_s)]}{[k_s^2 + (k_3 - p_3)^2 + m^2]}\CR
 a_s &=& -\frac{k_s f_0(k_s)}{k^2 + m^2}  \qquad \ \ \ \ \ \
 b_s = -\frac{k_s f_0(k_s)}{k_s^2 + (k_3 - p_3)^2 + m^2}  \CR
 a_3 &=& \frac{i k_3}{k^2 + m^2}  \qquad \ \ \ \ \ \ \ \
 b_3 = \frac{i (k_3 - p_3)}{k_s^2 + (k_3 - p_3)^2 + m^2}  \CR
\eea

The integral equations become
\bea\label{inteq}
\frac{\pi^2}{\lambda}\phi_s (k,p) &=& -\frac{k^+[p_3 + 2i k_s f_0(k_s)]}{[k^2+m^2][k_s^2+(k_3 - p_3)^2+m^2]}\int\frac{d^3q}{(k^+ -q^+)}\phi_s(q,p)\CR
&+&  \frac{k_s^2 + k_3(k_3-p_3)-m^2}{[k^2+m^2][k_s^2+(k_3 - p_3)^2+m^2]}\int\frac{d^3q}{(k^+ -q^+)}\phi_-(q,p)\CR
\frac{\pi^2}{\lambda}\phi_- (k,p) &=& -\frac{(k^+)^2}{[k^2+m^2][k_s^2+(k_3 - p_3)^2+m^2]}\int\frac{d^3q}{(k^+ -q^+)}\phi_s(q,p)\CR
&+&\frac{k^+[2i k_s f_0(k_s)-p_3]}{[k^2+m^2][k_s^2+(k_3 - p_3)^2+m^2]}\int\frac{d^3q}{(k^+ -q^+)}\phi_-(q,p)\eea

%It is easy to realize that for the case of static partons namely $k^+=k^-=0$, $a^+=a^-=b^+=b^-=0$ and hence  the only solution to the %equation is the trivial one.
We can now perform an integration over $k_3$ on both sides. Note that the integrals on the right-hand-sides do not depend on $k_3$.
 Thus the integration can be done directly, yielding an equation for $\tilde \phi(\tilde k,p) =\int dk_3 \phi(k,p)$ with $\tilde k = ( k_+,k_-)$.
% noting that  $\int d^3 q$ is independent of $k_3$. This will result in integral
%equations for $\tilde \phi$ on both sides.

To find the bound states, we have to go to Minkowski space, by analytic continuation to $p_3=i M_b$.
The solutions of the integral equation should provide us with the masses of the bound states $M_b$.

To perform the integrals over $k_3$, we will make use of the following   integrals
\bea
&&\int \frac{dk_3}{[k_s^2 + k_3^2 + m^2][k_s^2 + (k_3 - p_3)^2 + m^2]} = \frac{2\pi}{\sqrt{k_s^2 + m^2}} \frac {1}{[p_3^2 + 4(k_s^2 + m^2)]} \CR
&&\int dk_3\frac{k_3(k_3-p_3)}{[k_s^2 + k_3^2 + m^2][k_s^2 + (k_3 - p_3)^2 + m^2]} = \frac{2\pi\sqrt{k_s^2 + m^2}}{[p_3^2 + 4(k_s^2 + m^2)]}
\eea

The integral equations, after the $k_3$ integration, become
\bea\label{intequk3}
& & \frac{\pi}{2\lambda k^+} \sqrt{k_s^2 + m^2} [p_3^2 + 4(k_s^2 + m^2)]\tilde \phi_s (\tilde k,p_3) = -[2i k_s f_0(k_s)+p_3]\int\frac{d^2\tilde q}{(k^+ -q^+)}
\tilde\phi_s(\tilde q,p_3)\CR
&+&  4k^-\int\frac{d^2\tilde q}{(k^+ -q^+)}\tilde \phi_-(\tilde q,p_3)\CR
& & \frac{\pi}{2\lambda k^+} \sqrt{k_s^2 + m^2} [p_3^2 + 4(k_s^2 + m^2)]\tilde \phi_- (\tilde k,p_3) = -k^+\int\frac{d^2 \tilde q}{(k^+ -q^+)}\tilde \phi_s(\tilde q,p_3)\CR
&+&[2i k_s f_0(k_s)-p_3]\int\frac{d^2\tilde q}{(k^+ -q^+)}\tilde \phi_-(\tilde q,p_3)\CR\eea

% The wave functions above may have gauge dependent parts in them.
% However, we will be looking at the masses of the bound states, which depend on the gauge invariant parts only.
Note that, in the integrals on the right-hand-sides, what appears is  $\int\frac{dq^+}{k^+-q^+}\int dq^- \tilde\phi( \tilde q, p)$. The integration over $dq^-$ indicates that it is actually $x_- = 0$, and hence no need to have an integral of the form $i\int_0^{x_-} A_+(y) d y_-$ in the Wilson line between $\psi$ and $\bar\psi$.

It is amusing to note, that although it is $x_- = 0$ on the right-hand-sides, as follows from the integration $dq^-$, there is no such action on the left-hand-side.
However, if there is a bound state, it definitely has to appear in these equations, as it must couple to quark anti-quark combination.

\section{No massive bound states}\label{nbs}

Let us examine the equations (\ref{intequk3}), looking for possible solutions, namely bound states.

To this end, let us first examine the behavior at large $k$ components. This gives,

\bea\label{large k}
& & \frac{\pi}{\lambda}{k_s^2}\tilde \phi_s (\tilde k,p_3) \sim -i \lambda\int{d^2\tilde q}
\tilde\phi_s(\tilde q,p_3)
+  \frac {2k^-}{\sqrt k_s^2}\int{d^2\tilde q}\tilde \phi_-(\tilde q,p_3)\CR
& & \frac{\pi}{\lambda} {k_s^2} \tilde \phi_- (\tilde k,p_3) \sim -\frac{k^+}{2\sqrt k_s^2}\int{d^2 \tilde q}\tilde \phi_s(\tilde q,p_3)
+i \lambda\int{d^2\tilde q}\tilde \phi_-(\tilde q,p_3)\CR
& & k_1 , k_2 \rightarrow \infty \CR \eea

We have used also equation (\ref{fg}).
One may wonder whether
the region of $q^+$ near $k^+$ in the integrals of the right-had sides in (\ref{intequk3}) can  change the asymptotic behavior of $k^+$ due to a possible singularity. In fact there is no such a singularity. The proof of the absence of singularity is as follows.
%We would like to analyze the region of $q^+$ near $k^+$, on the right-hand-sides of (\ref{intequk3}), whether it contributes to the behavior for %large $k^+$,
%in view of the possible singularity.
%To this effect,
Consider the integral
\be
G_+(\tilde k,...)=\int\frac{d^2 \tilde q}{(k^+ -q^+)} F(\tilde q,...)
\ee
where $F(\tilde q,...)$ is a non-singular scalar function and the ... represent other variables.

We would like to regularize the denominator, by
\be
\frac{1}{(k^+ - q^+)} \rightarrow \frac{(k^- - q^-)}{[(\tilde k - \tilde q))^2 + \epsilon^2]}
\ee
We now get, for the function $G_+$, that
\be
G_+(\tilde k,...)=\int{d^2 \tilde q} \frac{(k^- - q^-)}{[(\tilde k - \tilde q))^2 + \epsilon^2]} F(\tilde q,...)=-\int{d^2 \tilde q} \frac{(q^-)}{[(\tilde q)^2 + \epsilon^2]} F(\tilde q + \tilde k,...)
\ee
For a small neighborhood $\delta$ around the origin in the last integral, we have
\be
\int_{|\tilde q| \leqslant \delta} {d^2 \tilde q} \frac{(q^-)}{[(\tilde q)^2 + \epsilon^2]} = 0
\ee
Thus there is  no singular contribution.

Now, from equation (\ref{phi}) it follows that
\bea\label{constants}
& & \int{d^2\tilde q}\tilde \phi_s(\tilde q,p_3) = - <meson(p)| \bar\psi(0) \psi(0)|0> \CR
& & \int{d^2\tilde q}\tilde \phi_-(\tilde q,p_3) = - <meson(p)| \bar\psi(0)\gamma_- \psi(0)|0> \CR \eea

Assuming a scalar bound state exists, we get that the right hand side of the first of equations (\ref{constants}) is non-zero,
while that of the second is zero.
\footnote {Note that we are in the rest frame of the bound state, so no $\_$ component for the matrix element of the current.}
This in turn implies that $\tilde \phi_s (\tilde k,p_3)$ behaves like $\frac{1}{k_s^2}$, as follows from the first of equations (\ref{large k}).
But then the integral on the left hand side of the first of equations (\ref{constants}) diverges.
Hence a contradiction.

Similarly, we can rule out vector bound states as well.

\section{ Bound-states in the bosonic theory}
Similar to the theory of a fundamental fermion coupled to a CS theory given in (\ref{actionfermion}) one can formulate a theory of a scalar in the fundamental representation coupled to a CS theory of the following form
\be\label{actionbosonic}
S_B= \frac{iK_B}{4\pi}\int d^3x Tr [A d A + \frac{2}{3} A^3] + D_\mu\bar \phi D^\mu \phi + m_B^2\bar\phi \phi + \frac{1}{2N_B}b_4 (\bar\phi \phi)^2
\ee
where now one defines $\lambda_B=\frac{N_B}{K_B}$. It was further shown in \cite{Aharony:2012ns} that in the Wilson - Fisher limit of
\be\label{bosonization}
b_4\rightarrow\infty,\qquad m_B\rightarrow \infty,\qquad \frac{4\pi m_B^2}{b_4} = |c_B|\ fixed
\ee
where $c_B^2$ is the $mass^2$ that appears in the full scalar propagator,
the bosonic theory  (\ref{actionbosonic}) is equivalent to the fermionic one
 (\ref{actionfermion}).
In \cite{Jain:2014nza} the $(2\rightarrow 2)$ scattering amplitude of the bosonic theory was  evaluated.  In particular it was found that for the range
$$-2\pi\lambda_B c_B (4-\lambda_B)\geq b_4\geq -2\pi c_B(4-\lambda_B^2)$$ the scattering amplitude admits poles which correspond to particle anti-particle bound states in the singlet channel.
Note that as the correspondence with the fermionic theory is only for $b_4$ tending to $\infty$, there will be no bound states in the fermionic case.

Furthermore, note that there are bound states in the bosonic theory also when there is no coupling to the gauge fields, namely in the limit of $\lambda_B = 0$.
The bound states are generated by the coupling $(\bar\phi \phi)^2$ alone in this case. Of course the coupling to gauge fields has an effect, as the relation among the parameters for
the appearance of bound states now involves $\lambda_B$ as well.

%%%%%%%%%%%%%%%%%%%%%%%%%%%%%%%%%
\section { Confronting the results of  \cite{Bardeen:2014qua}}

As mentioned in previous section, bound states in a Chern-Simons theory with scalars were considered \cite{Jain:2014nza}.
They find bound states for certain values of the parameters. They also find duality relations with the fermionic theory.
It turns out that the region of the scalar case, which is dual to the fermion case, has no bound states.
Our results are in accordance with those.

On the other hand, a massless scalar bound state for the massive fermion case was claimed in \cite{Bardeen:2014qua}, in contradiction with \cite{Jain:2014nza}.
Note that in \cite{Bardeen:2014qua} the relation $m_{bare} =  m (\frac{1}{\lambda}-\lambda)$ is employed, due to some assumed non-perturbative effect, and not (\ref{mbare}).
The latter, however, is the one derived in \cite{Frishman:2013dvg}.

One may adopt the more general point of view of two parameters, $m_{bare}$ and full mass $m$, with not necessarily a relation between them, as in \cite{Bardeen:2014qua}.
The origin of two mass parameters  according to \cite{Bardeen:2014qua} comes from a contribution of a  zero mode .

Adopting the two parameter approach, our relations (\ref{fg}) will have to change.
The function $f_0$ is now
\be
p_s f_0 = \lambda \sqrt{p_s^2 + m^2} + m_{bare}
\ee
Actually as in (\ref{initial f}). The function $g_0$ will change too, but the relation
\be
g_0 + f_0^2 = \frac {m^2}{p_s^2}
\ee
still holds, as in (\ref{fgrelation}).

Going over to the bound state equations, the relations (\ref{intequk3}) still hold, now with the two parameter $f_0$.
But the proof of no bound states still goes through, as the asymptotic behavior of $f_0$ for large $p_s$ is unchanged.

So we get no bound states also in the two parameter case, unlike \cite{Bardeen:2014qua}
\section{Summary and open questions}
In this note we have shown that the theory of fermions in the fundamental representation of $U(N)$ gauge symmetry  coupled to a CS term at level $K$ in the large N large K limits such that $\lambda= \frac{N}{K}$ is fixed does not admit fermion anti-fermion bound states. This provides a stronger statement than the one in our previous publication \cite{Frishman:2013dvg} where we have shown that high mass bound states  are forbidden. We derive this result by analyzing a 't Hooft like equation for the bound state wave-function.  We showed that assuming a non-vanishing wave-function we get a contradiction in the limit of large ``parton " momentum.

There are several open questions associated with results of this note:
\begin{itemize}
\item
Analyzing 't Hooft like equations for the wave-functions of bound states for the theory with scalars in the fundamental representation instead of fermions. As was discussed in section \S 6 in the theory of scalars due to the scalar self coupling there are bound states. It is interesting to determine the spectrum of masses of the bound states using the technique employed in this note

\item
One can generalize the case discussed in this note to the case of several fermion multiplets with several different masses. The naive intuition is that also for those cases there are no ferimon anti-fermion bound states. This statement could be verified by explicit generalization of the statement of this note.

\item
One can investigate the question of what are the necessary condition for an interaction of fundamental fermions to admit a spectrum of bound states. It is well known that coupling of the fermions to a Yang Mills term would yield bound states. The question is what other type of interactions share the same property.
\item
A very natural question is what of all the results of the exact solution of the theory of fermions coupled to a CS term in 3d are relevant to theories of non-abelian  fermions in four  space-time dimensions.
\end{itemize}

{\bf Acknowledgements}

We would like to thank Ofer Aharony for his comments on the manuscript and for insightful conversations, and
Guy Gur-Ari for useful discussions. We would  also like  to thank   the correspondence with W. Bardeen.
J. S. acknowledges support of the ``Einstein Center of Theoretical Physics " at the Weizmann Institute.


\vskip 0.5cm


\begin{thebibliography}{999}

%\cite{Frishman:2013dvg}
\bibitem{Frishman:2013dvg}
  Y.~Frishman and J.~Sonnenschein,
  ``Breaking conformal invariance - Large N Chern-Simons theory coupled to massive fundamental fermions,''  JHEP {\bf 1312}, 091 (2013)  [arXiv:1306.6465 [hep-th]].  %%CITATION = ARXIV:1306.6465;%%

%\cite{Giombi:2011kc}
\bibitem{Giombi:2011kc}
  S.~Giombi, S.~Minwalla, S.~Prakash, S.~P.~Trivedi, S.~R.~Wadia and X.~Yin,
  ``Chern-Simons Theory with Vector Fermion Matter,''
  Eur.\ Phys.\ J.\ C {\bf 72} (2012) 2112
  [arXiv:1110.4386 [hep-th]].
  %%CITATION = ARXIV:1110.4386;%%

%\cite{Aharony:2012ns}
\bibitem{Aharony:2012ns}
  O.~Aharony, S.~Giombi, G.~Gur-Ari, J.~Maldacena and R.~Yacoby,
  ``The Thermal Free Energy in Large N Chern-Simons-Matter Theories,''
  JHEP {\bf 1303}, 121 (2013)
  [arXiv:1211.4843 [hep-th]].
  %%CITATION = ARXIV:1211.4843;%%
  %5 citations counted in INSPIRE as of 12 Apr 2013

%\cite{Aharony:2012nh}
\bibitem{Aharony:2012nh}
  O.~Aharony, G.~Gur-Ari and R.~Yacoby,
  ``Correlation Functions of Large N Chern-Simons-Matter Theories and Bosonization in Three Dimensions,''
  JHEP {\bf 1212}, 028 (2012)
  [arXiv:1207.4593 [hep-th]].
  %%CITATION = ARXIV:1207.4593;%%
  %16 citations counted in INSPIRE as of 12 Apr 2013

%\cite{Aharony:2011jz}
\bibitem{Aharony:2011jz}
  O.~Aharony, G.~Gur-Ari and R.~Yacoby,
  ``d=3 Bosonic Vector Models Coupled to Chern-Simons Gauge Theories,''
  JHEP {\bf 1203}, 037 (2012)
  [arXiv:1110.4382 [hep-th]].
  %%CITATION = ARXIV:1110.4382;%%
  %34 citations counted in INSPIRE as of 12 Apr 2013

%\cite{GurAri:2012is}
\bibitem{GurAri:2012is}
  G.~Gur-Ari and R.~Yacoby,
  ``Correlators of Large N Fermionic Chern-Simons Vector Models,''
  JHEP {\bf 1302}, 150 (2013)
  [arXiv:1211.1866 [hep-th]].
  %%CITATION = ARXIV:1211.1866;%%
  %2 citations counted in INSPIRE as of 12 Apr 2013

%\cite{Maldacena:2012sf}
\bibitem{Maldacena:2012sf}
  J.~Maldacena and A.~Zhiboedov,
  ``Constraining conformal field theories with a slightly broken higher spin symmetry,''
  arXiv:1204.3882 [hep-th].
  %%CITATION = ARXIV:1204.3882;%%
  %40 citations counted in INSPIRE as of 12 Apr 2013

%\cite{Maldacena:2011jn}
\bibitem{Maldacena:2011jn}
  J.~Maldacena and A.~Zhiboedov,
   ``Constraining Conformal Field Theories with A Higher Spin Symmetry,''
  arXiv:1112.1016 [hep-th].
  %%CITATION = ARXIV:1112.1016;%%
  %52 citations counted in INSPIRE as of 12 Apr 2013

%\cite{Giombi:2011ya}
\bibitem{Giombi:2011ya}
  S.~Giombi and X.~Yin,
  ``On Higher Spin Gauge Theory and the Critical O(N) Model,''
  Phys.\ Rev.\ D {\bf 85} (2012) 086005
  [arXiv:1105.4011 [hep-th]].
  %%CITATION = ARXIV:1105.4011;%%
  %47 citations counted in INSPIRE as of 12 Apr 2013

%\cite{Giombi:2009wh}
\bibitem{Giombi:2009wh}
  S.~Giombi and X.~Yin,
  ``Higher Spin Gauge Theory and Holography: The Three-Point Functions,''
  JHEP {\bf 1009} (2010) 115
  [arXiv:0912.3462 [hep-th]].
  %%CITATION = ARXIV:0912.3462;%%
  %111 citations counted in INSPIRE as of 22 May 2013
%\cite{Takimi:2013zca}
\bibitem{Takimi:2013zca}
  T.~Takimi,
  ``Duality and Higher Temperature Phases of Large $N$ Chern-Simons Matter Theories on $S^2 \times S^1$,''  arXiv:1304.3725 [hep-th].  %%CITATION = ARXIV:1304.3725;%%


%\cite{Jain:2013py}
\bibitem{Jain:2013py}
  S.~Jain, S.~Minwalla, T.~Sharma, T.~Takimi, S.~R.~Wadia and S.~Yokoyama,
  ``Phases of large $N$ vector Chern-Simons theories on $S^2 \times S^1$,''  arXiv:1301.6169 [hep-th].  %%CITATION = ARXIV:1301.6169;%%  %4 citations counted in INSPIRE as of 19 May 2013

%\cite{Yokoyama:2012fa}
\bibitem{Yokoyama:2012fa}
  S.~Yokoyama,
  ``Chern-Simons-Fermion Vector Model with Chemical Potential,''  JHEP {\bf 1301} (2013) 052  [arXiv:1210.4109 [hep-th]].  %%CITATION = ARXIV:1210.4109;%%  %3 citations counted in INSPIRE as of 19 May 2013

%\cite{Giombi:2012ms}
\bibitem{Giombi:2012ms}
  S.~Giombi and X.~Yin,
  ``The Higher Spin/Vector Model Duality,''  arXiv:1208.4036 [hep-th].  %%CITATION = ARXIV:1208.4036;%%  %28 citations counted in INSPIRE as of 19 May 2013
%\cite{Banerjee:2012gh}
\bibitem{Banerjee:2012gh}
  S.~Banerjee, S.~Hellerman, J.~Maltz and S.~H.~Shenker,
  ``Light States in Chern-Simons Theory Coupled to Fundamental Matter,''  JHEP {\bf 1303} (2013) 097  [arXiv:1207.4195 [hep-th]].  %%CITATION = ARXIV:1207.4195;%%  %15 citations counted in INSPIRE as of 19 May 2013

%\cite{'tHooft:1974hx}
\bibitem{'tHooft:1974hx}
  G.~'t Hooft,
  ``A Two-Dimensional Model for Mesons,''
  Nucl.\ Phys.\ B {\bf 75} (1974) 461.
  %%CITATION = NUPHA,B75,461;%%
  %1529 citations counted in INSPIRE as of 09 May 2013

\bibitem{FSbook}
An elaborated review can be found in the book   Y. Frishman and J. Sonnenschein ``Non-perturbative Field Theory, from two dimensional conformal field theory to QCD in four dimensions" Cambridge monographs on mathematical physics. 
%\cite{Jain:2014nza}
\bibitem{Jain:2014nza}
  S.~Jain, M.~Mandlik, S.~Minwalla, T.~Takimi, S.~R.~Wadia and S.~Yokoyama,
  ``Unitarity, Crossing Symmetry and Duality of the S-matrix in large N Chern-Simons theories with fundamental matter,''
  arXiv:1404.6373 [hep-th].
  %%CITATION = ARXIV:1404.6373;%%
  %1 citations counted in INSPIRE as of 05 Jun 2014

%\cite{Bardeen:2014qua}
\bibitem{Bardeen:2014qua}
  W.~A.~Bardeen,
  ``The Massive Fermion Phase for the U(N) Chern-Simons Gauge Theory in D=3 at Large N,''
  %Submitted to: JHEP
  [arXiv:1404.7477 [hep-th]].
  %%CITATION = ARXIV:1404.7477;%%


\end{thebibliography}
\end{document}